%% file: SPAWC.tex
\documentclass[letter,10pt]{IEEEtran}
\renewcommand{\baselinestretch}{0.99}

\usepackage{pgfplots}
\usepackage{hyperref}
\usetikzlibrary{shapes.multipart,intersections}
\usepackage{cite}
\usepackage{amsmath,amssymb,amsfonts,steinmetz}
\usepackage{algorithmic}
\usepackage{graphicx}
\usepackage{textcomp}
\usepackage{xcolor}
\def\BibTeX{{\rm B\kern-.05em{\sc i\kern-.025em b}\kern-.08em
    T\kern-.1667em\lower.7ex\hbox{E}\kern-.125emX}}

\usepackage{tikz}
\usetikzlibrary{calc}
\makeatletter
\newcommand{\gettikzxy}[3]{%
  \tikz@scan@one@point\pgfutil@firstofone#1\relax
  \edef#2{\the\pgf@x}%
  \edef#3{\the\pgf@y}%
}
\makeatother

\usepackage[T1]{fontenc}
\usepackage[latin9]{inputenc}
\usepackage{bm}
\usepackage{amsmath}
\usepackage{subcaption}
\usepackage{amssymb}
\usepackage{stmaryrd}
\usepackage{babel}
\usepackage{comment}
\usepackage{graphics, graphicx}
\usepackage{algorithm,algorithmic}
\usepackage{xcolor}
\usepackage{gensymb}
\usepackage{cite}
\usepackage{enumitem}
\usepackage{url}

\usepackage{amsthm}
\input{./Definitions.tex}

\DeclareMathAlphabet\mathbfcal{OMS}{cmsy}{b}{n}

\begin{document}
%\title{Harnessing Extremely Large Antenna Arrays for ISAC: A Region-Wide Sensing Approach}
\title{DMA Reception for Simultaneous Area-Wide Sensing and Multi-User Uplink Communications}
\author{\IEEEauthorblockN{
Ioannis Gavras and George C. Alexandropoulos 
} \vspace{-0.8cm}
%\\
%\IEEEauthorblockA{Department of Informatics and Telecommunications, National and Kapodistrian University of Athens\\Panepistimiopolis Ilissia, 16122 Athens, Greece}
%\\
%\IEEEauthorblockA{emails: \{giannisgav, alexandg\}@di.uoa.gr}\vspace{-0.4cm}
}

\maketitle

\begin{abstract}
The recent surge in deploying extremely large antenna arrays is expected to play a vital role in future sixth generation wireless networks, enabling advanced radar target localization with enhanced angular and range resolution. This paper focuses on the promising technology of Dynamic Metasurface Antennas (DMAs), integrating numerous sub-wavelength-spaced metamaterials within a single aperture, and presents a novel framework for designing its analog reception beamforming weights with the goal to optimize sensing performance within a spatial Area of Interest (AoI), while simultaneously guaranteeing desired multi-user uplink communication performance. We derive the Cram\'{e}r-Rao Bound (CRB) with DMA-based reception for both passive and active radar targets lying inside the AoI, which is then used as the optimization objective for configuring the discrete tunable phases of the metamaterials. Capitalizing on the DMA partially-connected architecture, we formulate the design problem as convex optimization and present both direct CRB minimization approaches and low complexity alternatives using a lower-bound  approximation. Simulation results across various scenarios validate the effectiveness of the proposed framework, showing it consistently outperforms existing state-of-the-art methods.
\end{abstract}

\begin{IEEEkeywords}
Dynamic metasurfaces antennas, beamforming, Cram\'{e}r-Rao bound, localization, near-field, area-wide sensing.
\end{IEEEkeywords}
\let\thefootnote\relax\footnotetext{The authors are with the Dpt. of Informatics and Telecommunications, National and Kapodistrian University of Athens, Panepistimiopolis Ilissia, 16122 Athens, Greece (emails: \{giannisgav, alexandg\}@di.uoa.gr). This work has been supported by the SNS JU projects TERRAMETA and 6G-DISAC under the EU's Horizon Europe research and innovation program under Grant Agreement numbers 101097101 and 101139130, respectively.}

\vspace{-0.2cm}
\section{Introduction}
EXtremely Large (XL) antenna arrays are anticipated to be widely deployed in next generation wireless networks~\cite{XLMIMO_tutorial,41,hua2024near}, enabling operation primarily in the challenging near-field region, while providing numerous spatial degrees of freedom for communications, localization, and sensing applications~\cite{6G-DISAC_mag}. Dynamic Metasurface Antennas (DMAs) constitute a recent power- and cost-efficient hybrid analog and digital transceiver architecture utilizing arbitrary large numbers of metamaterials with tunable responses, which are grouped into disjoint microstrips each attached to a Radio Frequency (RF) chain, being capable of realizing either analog transmit or analog Receive (RX) BeamForming (BF)~\cite{Shlezinger2021Dynamic,10505154}.  

The beam focusing capability of a DMA-based transmitter serving multiple Users' Equipment (UEs) within its near-field regime was firstly investigated in~\cite{zhang2022beam}. An autoregressive attention neural network for non-line-of-sight user tracking using an RX DMA was proposed in~\cite{Nlos_DMA}, while \cite{NF_beam_tracking} introduced a near-field beam tracking framework for the same XL Multiple-Input Multiple-Output (MIMO) architecture. Additionally, in~\cite{FD_HMIMO_2023,spawc2024}, efficient designs for the digital and analog BF matrices of DMA-based full duplex transceivers were presented targeting near-field Integrated Sensing And Communications (ISAC). The near-field localization capabilities of an RX DMA were studied in~\cite{gavras2025near}, optimizing the analog RX BF weights using the Cram\'{e}r-Rao Bound (CRB) for multi-target localization. Very recently, in~\cite{XL_MIMO_ISAC}, a DMA-based in band full duplex XL MIMO system was optimized for simultaneous communications and monostatic-type sensing. However, to the best authors' knowledge, most of the research on XL MIMO systems for ISAC focuses on BF optimization for localization instead of a sensing area illumination. Interestingly, the additional degrees of freedom present in XL MIMO can  be leveraged for broader area-wide sensing in conjunction with communications, allowing for simultaneous high-resolution radar target localization within a predefined Area of Interest (AoI).

In this paper, we study the optimization of DMA-based reception for area-wide sensing within a desired AoI, while simultaneously supporting Quality-of-Sevice (QoS) constraints for multi-user uplink communications. We first derive the CRB and calculate the corresponding Position Error Bound (PEB) for both active users and passive targets lying within the AoI, using the former metric as the minimization objective for designing the DMA's analog RX BF weights. By exploiting the partially-connected analog BF architecture of the considered RX DMA, we reformulate the design objective as a convex optimization problem and propose optimization solutions to minimize the CRB across the AoI, along with low complexity alternatives based on a CRB lower bound approximation. Our simulation results validate the effectiveness of the proposed framework for AoI-wide sensing, showcasing the superiority of the proposed dual-functional RX DMA design framework over conventional approaches.

\textit{Notations:}
Vectors and matrices are represented by boldface lowercase and uppercase letters, respectively. The transpose, Hermitian transpose, inverse, and Euclidean norm are denoted as $(\cdot)^{\rm T}$, $(\cdot)^{\rm H}$, $(\cdot)^{-1}$, and $\|\cdot\|$, respectively. $[\mathbf{A}]_{i,j}$ and $[\mathbf{A}]_{i:j,u:v}$ give respectively $\mathbf{A}$'s $(i,j)$th element and sub-matrix spanning rows $i$ to $j$ and columns $u$ to $v$. $\mathbf{I}_{n}$ and $\mathbf{0}_{n}$ ($n\geq2$) are the $n\times n$ identity and zeros' matrices, respectively, and $\boldsymbol{1}_{N}$ is an $N \times 1$ column vector of ones. $\mathbb{C}$ is the complex number set, $|a|$ is the amplitude of scalar $a$, and $\jmath\triangleq\sqrt{-1}$. $\mathbb{E}\{\cdot\}$, ${\rm Tr}\{\cdot\}$, and $\Re\{\cdot\}$ give the expectation, trace, and real part, respectively. $\mathbf{x}\sim\mathcal{CN}(\mathbf{a},\mathbf{A})$ indicates a complex Gaussian random vector with mean $\mathbf{a}$ and covariance matrix $\mathbf{A}$.

%\vspace{-0.1cm}
\section{System and Channel Models}
We consider a wireless system setup centered around an RX equipped with an XL DMA panel~\cite{Shlezinger2021Dynamic}, which wishes to communicate in the uplink direction with $U$ single-antenna UEs, while simultaneously being capable to sense $K\geq U$ radar targets (including the $U$ UEs and $K-U$ passive targets) within a sensing AoI. DMAs efficiently enable the integration of numerous sub-wavelength-spaced metamaterials, which are usually grouped in microstrips each attached to a reception RF chain (comprising a low noise amplifier, a mixer downconverting the received signal from RF to baseband, and an analog-to-digital converter), within possibly extremely large apertures~\cite{41}. By dynamically adjusting the response of metamaterials to the incoming signal, goal-oriented analog RX BF can be achieved. In this paper, we are interested for simultaneous AoI-wide sensing and multi-user uplink communications. The RX DMA panel is situated in the $xz$-plane with the first microstrip positioned at the origin. There exist in total $N_{\rm RF}$ microstrips within the panel, each composed of $N_{\rm E}$ distinct metamaterials placed in a uniform linear pattern with $d_{\rm E}$ distance between adjacent elements. The microstrips are individually linked to RX RF chains, which are separated from one another by a distance of $d_{\rm RF}$. Consequently, the RX DMA includes in total $N\triangleq N_{\rm RF}N_{\rm E}$ metamaterials. The position of each $k$th radar target is expressed by its spherical coordinates $(r_k,\theta_k,\varphi_k)$ $\forall k=1,\ldots,K$, including the distances from the origin, as well as the elevation and azimuth angles, respectively. From those targets, $U$ out of $K$ with coordinates $(r_u,\theta_u,\varphi_u)$ $\forall u=1,\ldots,U$ are the UEs.

We define the $N\times N$  diagonal matrix $\P_{\rm RX}$, whose elements model the signal propagation inside the RX DMA's microstrips $\forall$$i=1,\dots,N_{\rm RF}$ and $\forall$$n = 1,\dots,N_{\rm E}$~\cite{FD_HMIMO_2023}, as $[\P_{\rm RX}]_{((i-1)N_{\rm E}+n,(i-1)N_{\rm E}+n)} \triangleq \exp{(-\rho_{i,n}(\alpha_i + \jmath\beta_i))}$, where $\alpha_i$ is the waveguide attenuation coefficient, $\beta_i$ is the wavenumber, and $\rho_{i,n}$ denotes the location of the $n$th element in the $i$th microstrip (extension to lossy microstrips are left for future work~\cite{DMA_Losses}). Let also $w^{\rm RX}_{i,n}$ be the adjustable response (i.e., analog weight), associated with the $n$th metamaterial of each $i$th microstrip, which is assumed to conform to a Lorentzian-constrained phase model and is mapped to the phase profile codebook $\mathcal{W}$:
\begin{align}\label{eq: code}
    w^{\rm RX}_{i,n} \in \mathcal{W}\triangleq \{0.5\left(\jmath+e^{\jmath\phi}\right)|\phi\in\left[-\pi/2,\pi/2\right]\}.
\end{align}
Using this definition, the DMA's analog RX BF matrix $\W_{\rm RX}\in\mathbb{C}^{N\times N_{\rm RF}}$ is obtained as $[\W_{\rm RX}]_{((i-1)N_{\rm E}+n,j)}=w^{\rm RX}_{i,n}$, for $i=j$, and $[\W_{\rm RX}]_{((i-1)N_{\rm E}+n,j)}=0$, for $i\neq j$.

\subsection{Near-Field Channel Model}
We focus on the upper part of the FR3 frequency band~\cite{FR3} and above, where with the deployment of XL MIMO systems, near-field channel conditions will frequently take place~\cite{liu2023near}. To this end, we model the $N$-element complex-valued channel vector between the RX DMA and each $u$th single-antenna UE as $\h_u\triangleq\h_{{\rm UL},u}+\h_{{\rm R},u}$, where the first term, $\h_{{\rm UL},u}\in\Compl^{N\times 1}$, represents the Line-of-Sight (LoS) channel expressed as:
\begin{align}\label{eqn:UL_chan}
    [\h_{{\rm UL},u}]_{(i-1)N_{\rm E}+n} \triangleq \sqrt{\frac{{F(\theta_{u,i,n})}}{{(4\pi)^2 r_{u,i,n}^2}}} \exp\left(\frac{\jmath2\pi}{\lambda} r_{u,i,n}\right)
\end{align}
with $\lambda$ denoting the signal wavelength, $F(\cdot)$ represents each metamaterial's radiation profile, as well as $r_{u,i,n}$ and $\theta_{u,i,n}$ stand for the distance and elevation angle, respectively, from the 
$u$th UE's antenna to the $n$th metamaterial of the $i$th microstrip of the RX DMA panel; these channel parameters can be computed similar to~\cite{FD_HMIMO_2023}. The second term $\h_{{\rm R},u}\in\Compl^{N\times 1}$ models the scattered channel components between the RX DMA and the $u$th UE from the rest $K-1$ targets (i.e., the rest of the $U-1$ UEs and the $K-U$ passive targets), when considered as point sources, which can be expressed as: 
\begin{align}\label{eq: H_R}
    \h_{{\rm R},u}&\triangleq\sum_{\substack{k=1,k\neq u}}^{K}\beta_k\boldsymbol{\alpha}_{\rm RX}(r_k,\theta_k,\phi_k),%\alpha_{\rm TX}^{\rm H}(r_k,\theta_k,\phi_k),
\end{align}
where $\beta_k\in\Compl$ represents the reflection coefficient for each $k$th radar target and ${\boldsymbol{\alpha}}_{\rm RX}(r_k,\theta_k,\phi_k)$ denotes the complex-valued near-field RX steering vectors, which can be modeled analogously to \cite{FD_HMIMO_2023}.
%${\boldsymbol{\alpha}}_{\rm RX}(r_k,\theta_k,\phi_k)$ and $\alpha_{\rm TX}(r_k,\theta_k,\phi_k)$ denote the RX and TX complex-valued near-field steering vectors, which are modeled analogously to \cite{FD_HMIMO_2023}.

\subsection{Received Signal Model}
The baseband received signal at the outputs of the DMA's RX RF chains after $T$ UE data transmissions can be mathematically expressed via the matrix $\Y\triangleq [\y(1),\ldots,\y(T)]\in\Compl^{N_{\rm RF}\times T}$ with $t=1,\ldots,T$:
\begin{align}\label{eq:UL_signal_matrix}
    \Y = \W_{\rm RX}^{\rm H}\P_{\rm RX}^{\rm H}\sum_{u=1}^U\h_u\s_u+\N,
\end{align}
where $\N \triangleq [\n(1),\ldots,\n(T)]\in\Compl^{N\times T}$ with each $\n(t)\sim\mathcal{CN}(\mathbf{0},\sigma^2\mathbf{I}_{\rm N_{\rm RF}})$ being the Additive White Gaussian Noise (AWGN) vector, and $\s_u \triangleq [s_u(1),\ldots,s_u(T)]\in\Compl^{1\times T}$. Each transmitted communication signal $s_u(t)$ is subjected to the power constraint $\mathbb{E}\{\|s_u(t)\|^2\}\leq P_{\max}$, where $P_{\max}$ signifies the maximum UE transmission power in the uplink. 

We make the reasonable assumption that accurate estimates for $\h_u$ $\forall u$ are available per coherent channel block at the RX DMA through Sounding Reference Signals (SRSs) and any of the well-known channel estimation schemes. This enables the computation at the RX of the per $u$th UE instantaneous received Signal-to-Noise Ratio (SNR) value $\Gamma_u\triangleq\left\|\W_{\rm RX}^{\rm H}\P_{\rm RX}^{\rm H}\h_{u}\right\|^2/\sigma^2$ for a given $\P_{\rm RX}$ and any of the available $\W_{\rm RX}$ matrices. Note that, using SRSs as described in~\cite{5G_NR_positioning}, estimates for $(r_u,\theta_u,\varphi_u)$ $\forall u$ could be also obtained, and then used to compose the LoS channel components in~\eqref{eqn:UL_chan}.
%To this end, we assume that estimates $(\widehat{r}_k,\widehat{\theta}_k,\widehat{\varphi}_k)$, $\forall k$ of the radar targets spatial parameters are available per coherent channel block at the RX DMA through Sounding Reference Signals (SRSs)~\cite{5G_NR_positioning}, and can be used to generate an estimate for $\widehat{\h}_u$, $\forall u$. Note that the reflection coefficient $\beta_{\rm k}$, $\forall k$ included in $\h_u$ is unknown and cannot be estimated from those sounding signals.
%\vspace{-0.1cm}

\section{Proposed DMA Design for ISAC}\label{Sec: 3}
In this section, we present our RX DMA design for concurrent near-field AoI-wide sensing and uplink communications. Initially, we derive the CRB and compute the corresponding PEB for both the passive and active radar targets within the AoI. The latter metric is then used as the optimization objective in the proposed framework that seeks for the optimal dual-functional RX analog BF design.
%, where we present optimal BF optimization solutions alongside efficient, low-complexity closed-form alternatives for determining the RX DMA's analog BF weights.

\subsection{PEB Performance Analysis} %using \eqref{eq:UL_signal}
%Let $\mathbb{A}$ be the set of all the possible positions within the AoI. To this end, we discretize $\mathbb{A}$ into a finite set of points, each represented by the polar coordinates $(r_a, \theta_a, \phi_a)$ with $a=1,2,\ldots,A$.  Given the LoS UL channel $\h_{{\rm UL},a}$, $\forall a$ as defined in \eqref{eqn:UL_chan}, we assume that the received signals originate exclusively within $\mathbb{A}$. 

Let $\s_u$ $\forall u$ include the individual unit-powered uplink data streams for a coherent channel block involving $T$ symbol transmissions, implying that $T^{-1}\s_u\s_u^{\rm H}=1$. By inspecting, \eqref{eq:UL_signal_matrix}'s, it follows that the received signal at the output of the RX DMA can be modeled as $\y\triangleq{\rm vec}\{\Y\}\sim\mathcal{CN}(\boldsymbol{\mu},\sigma^2\I_{N_{\rm RF}T})$, with mean $\boldsymbol{\mu} \triangleq {\rm vec}\{\W_{\rm RX}^{\rm H}\P_{\rm RX}^{\rm H}\sum_{u=1}^U\h_{u}\s_u\}$. In the context of estimating $\boldsymbol{\xi}\triangleq[r_1,\theta_1,\phi_1,\ldots,r_K,\theta_K,\phi_K]^{\rm T}$, the $3K\times3K$ Fisher Information Matrix (FIM)\cite{kay1993fundamentals} is defined as follows:
\begin{align}\label{eq: FIM}
    \mathbfcal{I}_{\boldsymbol{\xi}} =\begin{bmatrix}
    \mathbfcal{I}_{r_1r_1} & \mathbfcal{I}_{r_1\theta_1} & \mathbfcal{I}_{r_1\phi_1} & \ldots & \mathbfcal{I}_{r_1\phi_K}\\
    \mathbfcal{I}_{\theta_1r_1} & \mathbfcal{I}_{\theta_1\theta_1} & \mathbfcal{I}_{\theta_1\phi_1} & \ldots &\mathbfcal{I}_{\theta_1\phi_K}   \\
    \mathbfcal{I}_{\phi_1r_1} & \mathbfcal{I}_{\phi_1\theta_1} & \mathbfcal{I}_{\phi_1\phi_1} & \ldots &\mathbfcal{I}_{\phi_1\phi_K}  \\
    \vdots & \vdots & \vdots & \ddots & \vdots  \\
    \mathbfcal{I}_{\phi_K r_1} & \mathbfcal{I}_{\phi_K\theta_1} & \mathbfcal{I}_{\phi_K\phi_1} & \ldots &\mathbfcal{I}_{\phi_K\phi_K} 
    \end{bmatrix}
\end{align}
with the $(i,j)$th element of the FIM (with $i,j\in\boldsymbol{\xi}$) given by 
\begin{align}
    [\mathbfcal{I}]_{i,j}\!=\! \frac{2}{\sigma^2}\Re\left\{\!\frac{\partial \boldsymbol{\mu}^{\rm H}}{\partial[\boldsymbol{\xi}]_i}\frac{\partial \boldsymbol{\mu}}{\partial[\boldsymbol{\xi}]_j}\!\right\}\!.
\end{align}
According to the vectorized received signal matrix from \eqref{eq:UL_signal_matrix}, the partial derivatives can be computed as follows $\forall i\in\boldsymbol{\xi}$:
\begin{align}\label{eq: deriv_mean}
    \frac{\partial\boldsymbol{\mu}}{\partial[\boldsymbol{\xi}]_i} = {\rm vec}\left\{\W_{\rm RX}^{\rm H}\P_{\rm RX}^{\rm H}\sum_{u=1}^U\frac{\partial\h_{u}}{\partial[\boldsymbol{\xi}]_i}\s_u\right\}.
\end{align}
Let $\Q_{\rm RX}\triangleq\W_{\rm RX}\W_{\rm RX}^{\rm H}$, then each element of the FIM matrix can be represented as:
\begin{align}
    \nonumber[\mathbfcal{I}]_{i,j} =\frac{2T}{\sigma^2}\Re\Bigg\{{\rm Tr}\left\{\P_{\rm RX}^{\rm H}\sum_{u=1}^{U}\frac{\partial\h_{u}}{\partial[\boldsymbol{\xi}]_j}\frac{\partial\h_{u}^{\rm H}}{\partial[\boldsymbol{\xi}]_i}\P_{\rm RX}\Q_{\rm RX}\right\}\Bigg\}.
\end{align}

Putting all above together, the PEB for the targets' positions $(r_1,\theta_1,\phi_1),\ldots,(r_K,\theta_K,\phi_K)$ can be expressed as a function of the DMA RX analog BF weights, as follows~\cite{kay1993fundamentals}:
\begin{align}\label{eq:PEB}
\text{PEB}(\W_{\rm RX};\boldsymbol{\xi}) \triangleq 
%\sqrt{{\rm Tr}\{\text{CRB}(r,\theta,\phi)\}}=
\sqrt{\text{CRB}(\W_{\rm RX};\boldsymbol{\xi})}=\sqrt{{\rm Tr}\left\{\mathbfcal{I}^{-1}_{\boldsymbol{\xi}}\right\}}.
\end{align}

%\vspace{-0.5cm}

\subsection{Dual-Functional RX DMA Design}
We focus on designing the analog BF weights $\W_{\rm RX}$ at the RX DMA so as to enable sensing in a desired AoI, while simultaneously ensuring compliance with a set of QoS constraints for multi-user uplink communications. We particularly define any desired AoI $\mathbb{A}$ as a finite set of $|\mathbb{A}|$ discrete positions: $(r_a, \theta_a, \phi_a)$ $\forall a = 1,\ldots,|\mathbb{A}|$ ($|\cdot|$ here indicates set cardinality). To this end, by formulating \eqref{eq:PEB} for $\mathbb{A}$'s discretization points $\boldsymbol{\eta}\triangleq[r_1,\theta_1,\phi_1,\ldots,r_{|\mathbb{A}|},\theta_{|\mathbb{A}|},\phi_{|\mathbb{A}|}]$, we seek for $\W_{\rm RX}$ solving the following optimization problem: 
\begin{align}
        \mathcal{P}_1: \nonumber\underset{\substack{\W_{\rm RX}}}{\min} \,\, {\rm PEB}(\W_{\rm RX};\boldsymbol{\eta})\,\,\,\text{\text{s}.\text{t}.}\,\,\,\Gamma_u\geq\gamma_u,\, w^{\rm RX}_{i,n} \in \mathcal{W}\,\,\forall u,i,n.
\end{align}
%where, the minimization objective is the PEB of the FIM as defined in \eqref{eq: FIM}, but evaluated for the positioning parameter vector $\boldsymbol{\eta}\triangleq[r_1,\theta_1,\phi_1,\ldots,r_A,\theta_A,\phi_A]$, which encompasses all positions within the AoI $\mathbb{A}$. 
The first constraint imposes the per $u$th UE SNR threshold $\gamma_u$, whereas the second constraint enforces all analog RX BF weights to conform to the Lorentzian structure in~\eqref{eq: code}. 

It can be easily concluded that $\mathcal{P}_1$ is in a non-convex form, due to the structure of the objective, but it can be relaxed via SemiDefinite Relexation (SDR). Specifically, by applying Schur's complement \cite{keskin2022optimal}, we introduce a set of auxiliary variables $\b\triangleq[b_1,b_2,\ldots,b_{|\mathbb{A}|}]\in\Compl^{3|\mathbb{A}|\times 1}$, which helps us to transform $\mathcal{P}_1$ to the following equivalent formulation:
\begin{align}
        &\nonumber\underset{\substack{\W_{\rm RX},\m}}{\min} \,\, \boldsymbol{1}_{3A}^{\rm T}\b
        \\&\nonumber\,\,\,\text{\text{s}.\text{t}.}\,\,\,\,{\rm Tr}\{\P_{\rm RX}^{\rm H}\h_{u}\h_{u}^{\rm H}\P_{\rm RX}\Q_{\rm RX}\}\geq\gamma_u\,\forall u,
        \\\nonumber&\qquad\,\, \begin{bmatrix}
        \boldsymbol{\mathcal{I}_{\boldsymbol{\eta}}} & \e_a\\
        \e_a^{\rm T} & b_a
        \end{bmatrix}\succeq0\, \forall a,\,w^{\rm RX}_{i,n} \in \mathcal{W}\,\,\forall i,n,
\end{align}  
where $\e_a$ is the $a$th column of the $3|\mathbb{A}|\times 3|\mathbb{A}|$ identity matrix. Note that the Lorentzian constraint is non-convex, but it can be straightforwardly incorporated into the optimization problem, as follows. Let $\w_i^{\rm RX}\triangleq[w_{i,1}^{\rm RX},w_{i,2}^{\rm RX},\ldots,w_{i,N_{\rm E}}^{\rm RX}]^{\rm T}\in\Compl^{N_{\rm E}\times1}$ include the adjustable response weights attached to each $i$th RF chain of the RX DMA. We construct the $N\times N$ diagonal block matrix $\Q_{\rm RX}$ with each $i$th block of size $N_{\rm E}\times N_{\rm E}$ having the structure $\w_i^{\rm RX}(\w_i^{\rm RX})^{\rm H}$. By using the Lorentzian constraint $\w_i^{\rm RX}=0.5(\jmath\boldsymbol{1}_{N_{\rm E}}+\q_i)$, with $\q_i\triangleq[q_{i,1},\ldots,q_{i,N_{\rm E}}]^{\rm T}\in\Compl^{N_{\rm E}\times1}$ (i.e., including unconstrained analog weights in the $\w_i^{\rm RX}$ expression), we can further simplify the latter optimization problem formulation as follows~\cite{gavriilidis2024metasurface}:
\begin{align}
        &\nonumber\underset{\substack{\{\Q_i\}_{i=1}^{N_{\rm RF}},\t}}{\min} \,\, \boldsymbol{1}_{3A}^{\rm T}\b
\\&\nonumber\,\,\,\text{\text{s}.\text{t}.}\,\,\,\sum_{i=1}^{N_{\rm RF}}{\rm Tr}\{\H_{u,i}\Q_i\}\geq\gamma_u\,\forall u,\,\Q_i\succeq 0\,\forall i, 
\\&\nonumber\,\,\qquad\begin{bmatrix}
        \boldsymbol{\mathcal{I}_{\boldsymbol{\eta}}} & \e_a\\
        \e_a^{\rm T} & b_a
        \end{bmatrix}\succeq0\, \forall a,
\end{align}
where $\H_{u,i},\Q_i\in\Compl^{(N_{\rm E}+1)\times(N_{\rm E}+1)}$ are defined as follows:
\begin{align}
&\nonumber\H_{u,i} \triangleq \begin{bmatrix}
\A_i&\A_i\jmath\boldsymbol{1}_{N_{\rm E}}\\
(\jmath\boldsymbol{1}_{N_{\rm E}})^{\rm H}\A_i & 0
\end{bmatrix},\,\,\Q_i\triangleq\begin{bmatrix}
\q_i \\
1
\end{bmatrix}
\begin{bmatrix}
\q_i^{\rm H}& 1
\end{bmatrix}
\end{align}
with $\A_i\triangleq[\H_u]_{i_1:iN_{\rm E},i_1:iN_{E}}$ and $\H_u=\P_{\rm RX}^{\rm H}\h_u\h_u^{\rm H}\P_{\rm RX}$. To solve this problem, we have applied a series of SDRs, where each rank-one constraint corresponding to $\Q_i$ $\forall i$ was replaced with a positive semidefinite one. Additionally, each element of $\mathbfcal{I}_{\boldsymbol{\eta}}$ has been reformulated in accordance with the block-diagonal structure of each $\Q_i$, similar to $\H_{u.i}$ $\forall u,i$. This transformation results in a convex problem, which can be efficiently solved using standard convex programming solvers (e.g., CVX~\cite{cvx}). Finally, the optimal solution $\mathbf{q}_i^{\rm opt}$ $\forall i$ for the given optimization problem can be obtained as $\mathbf{q}_i^{\rm opt} = \exp(\jmath\angle[\u_i]_{1:N_{\rm E}})$, with $\u_i$ representing the principal singular vector of the optimized $\Q_i$. It is noted that, as established in \cite{luo2010semidefinite}, the SDR-relaxed problem yields solutions that satisfy the necessary rank-one constraints.

\begin{figure*}[!t]
  \begin{subfigure}[t]{0.33\textwidth}
  \centering
    \includegraphics[width=\textwidth]{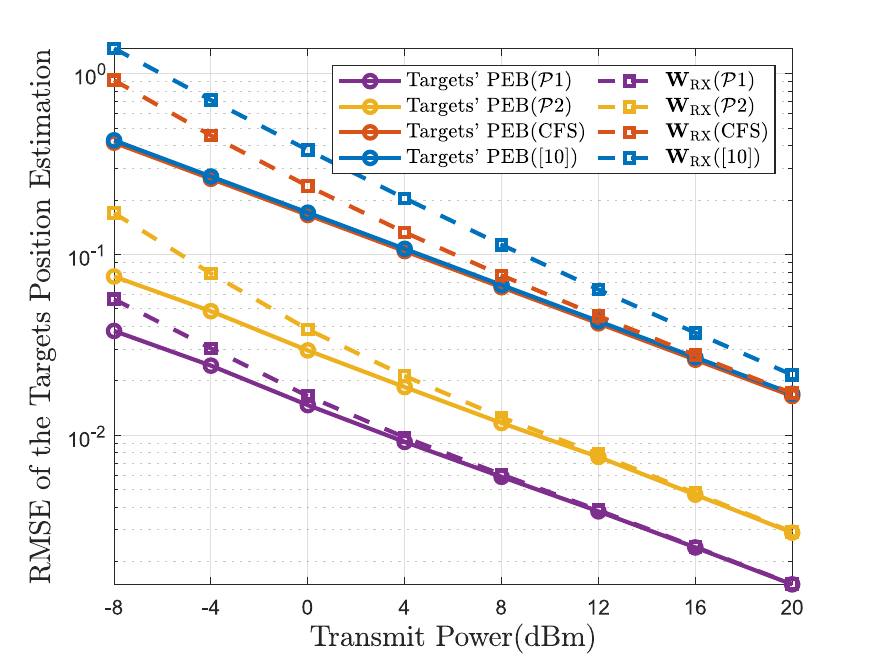}
    \caption{Sensing performance w.r.t. $P_{\rm{\max}}$ in dBm.}
    \label{fig: RMSE}
  \end{subfigure}\hfill
  \hfill
  \begin{subfigure}[t]{0.33\textwidth}
  \centering
    \includegraphics[width=\textwidth]{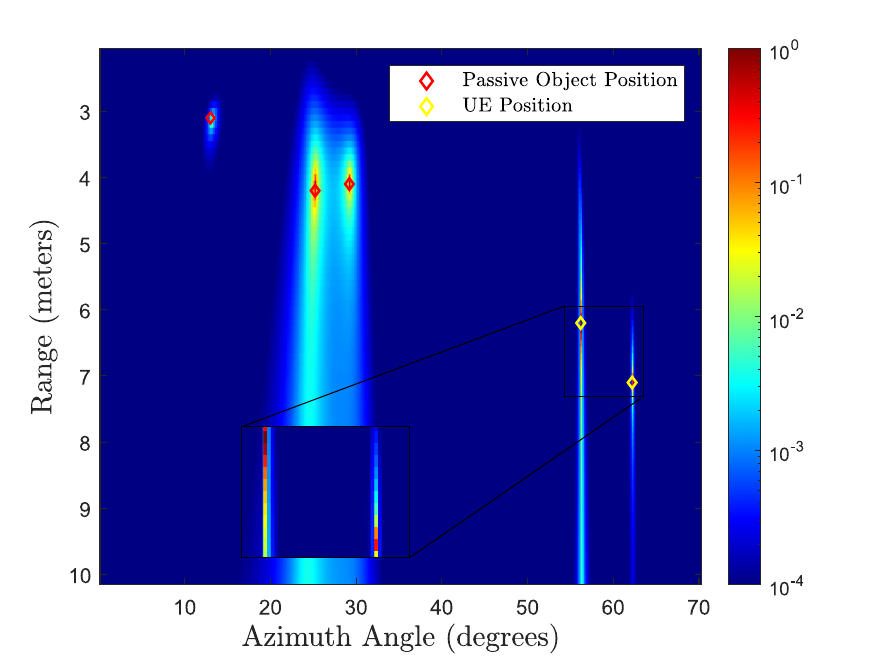}
    \caption{BF gain with the CFS solution.}
    \label{fig: BF}
  \end{subfigure}
  \begin{subfigure}[t]{0.33\textwidth}
  \centering
    \includegraphics[width=\textwidth]{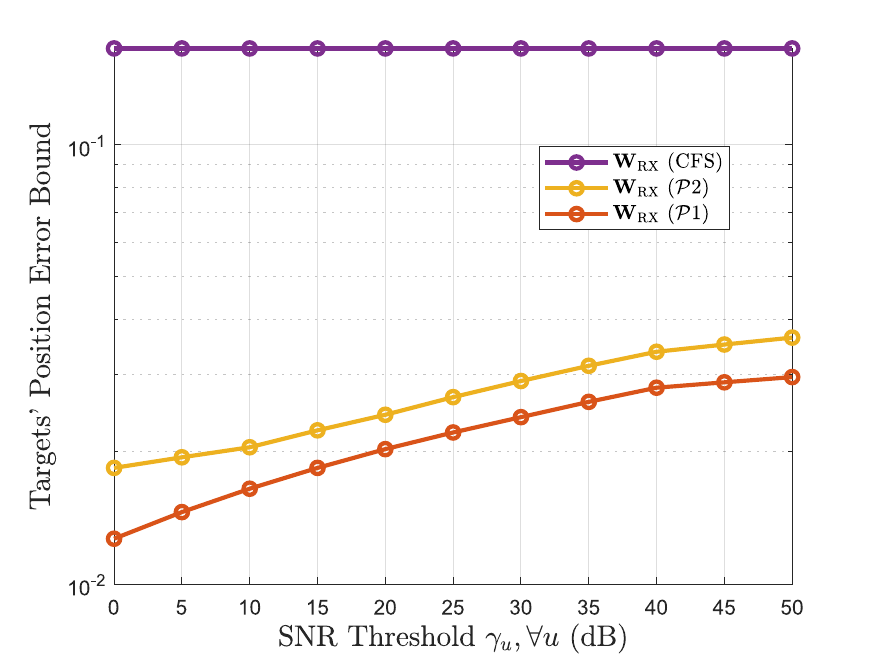}
    \caption{Trade-off for $P_{\rm{\max}}=0$ dBm.}
    \label{fig: SNR}
  \end{subfigure}
  \caption{\small{Sensing and communications performances with the proposed dual-functional RX DMA framework, considering a panel with $N_{\rm RF}=8$ microstrips, each hosting $N_{\rm E} = 64$ phase-tunable metamaterials.}}%\vspace{-0.4cm}
  \label{fig:Estimation_vs_P_T}
\end{figure*}
\subsection{Low Complexity Solutions}
It is evident that applying Schur's complement renders the problem convex, but introduces high-dimensional linear matrix inequalities, whose complexity scales with the number $\mathbb{A}$'s discretization points, $|\mathbb{A}|$, and the number of antenna elements in the RX DMA, $N$. To reduce this complexity, rather than directly optimizing the CRB, we can instead optimize its lower bound given by ${\rm Tr}\{\mathbfcal{I}^{-1}\}\geq{\rm Tr}\{\mathbfcal{I}\}^{-1}$ which follows from the harmonic-geometric mean inequality, as shown in~\cite{1}. Therefore, to indirectly minimize the PEB, we can maximize ${\rm Tr}\{\mathbfcal{I}\}$. This approach approximates the FIM as a block diagonal matrix, effectively ignoring off-diagonal elements and, consequently, the interdependence among the different points in $\mathbb{A}$. By applying the same simplifications as in $\mathcal{P}_1$, in particular, by incorporating the Lorentzian-codebook constraint into the optimization through a series of SDRs, we can formulate the following optimization problem for the intended dual-functional RX DMA design as follows:
\begin{align}
        &\mathcal{P}_2: \nonumber\underset{\substack{\{\Q_i\}_{i=1}^{N_{\rm RF}}}}{\max} \,\, \sum_{i=1}^{N_{\rm RF}}{\rm Tr}\{\B_i\Q_i\}
        \\&\nonumber\qquad\,\,\text{\text{s}.\text{t}.}\,\,\,\sum_{i=1}^{N_{\rm RF}}{\rm Tr}\{\H_{u,i}\Q_i\}\geq\gamma_u\,\forall u,\, \Q_i\succeq0\,\forall i,
\end{align}
where $\B_i\in\Compl^{(N_{\rm E}+1)\times(N_{\rm E}+1)}$ has a similar structure to $\H_{u,i}$, but is defined with respect to the  the matrix $\B\triangleq\sum_{a=1}^{3A}\sum_{i=1}^{3}\frac{\partial\h_{a}}{\partial[\boldsymbol{\xi}]_i}\frac{\partial\h_{a}^{\rm H}}{\partial[\boldsymbol{\xi}]_i}$. As a result, the $\mathcal{P}_2$ formulation is rendered convex allowing us to derive the optimum BF solution in the same manner as in $\mathcal{P}_1$. 

It is needs to be noted that, although $\mathcal{P}_2$ deals with the complexity scaling issue of $\mathcal{P}_1$ with respect to $|\mathbb{A}|$ value, computational challenges remain, particularly for XL MIMO arrays with very large $N$, especially when using SDR. To further simplify $\mathcal{P}_2$, we can exploit the fact that its optimal solution lies within the subspaces of both the communication and sensing channels. Specifically, if we disregard the multi-user uplink communication QoS constraints, the optimal solution for $\mathcal{P}_2$ can be determined by the principal singular vectors of the matrices $\B_i$ $\forall i$, implying that this solution resides solely in the sensing channel subspace. However, since $\mathbb{A}$ encompasses multiple points, the subspace of $\B$, and consequently $\B_i$ $\forall i$, can also approximate the communication channel subspace. Therefore, we can derive a Closed-Form Solution (CFS) that approximates the optimal solution of $\mathcal{P}_2$ via the principal singular vector of each $\B_i$ for each $\q_i$ $\forall i$ as $\mathbf{q}_i^{\rm approx} = \exp(\jmath\angle[\k_i]_{1:N_{\rm E}})$, where $\k_i$ represents the principal singular vector of $\B_i$.

\subsection{Complexity Analysis}
The computational complexities of $\mathcal{P}_1$ and $\mathcal{P}_2$ are primarily determined by the number of points $|\mathbb{A}|$ in the sensing AoI and the number $N$ of the RX DMA metamaterials. Both problems utilize interior-point methods for the SDRs, with a worst-case time complexity of $\mathcal{O}(n^2\sum_{i=1}^C m_i^2 + n\sum_{i=1}^C m_i^3)$~\cite{nemirovski2004interior}, where $n$ denotes the number of optimization variables and $C$ represents the number of linear matrix inequality constraints with and $m_i$ being the column or row size of each $i$th of these constraints. Specifically, for $\mathcal{P}_1$, we have $n = 3A + N_{\rm RF}(N_{\rm E}+1)^2$, $C = 3A + N_{\rm RF}$, and $m_i = N_{\rm E}+1$ for $1 \leq i \leq N_{\rm RF}$, while $m_i = 3A+1$ for $N_{\rm RF}+1 \leq i \leq C$. For $\mathcal{P}_2$, which involves only the $N_{\rm RF}$ matrices, the parameters are $n = N_{\rm RF}(N_{\rm E}+1)^2$, $C = N_{\rm RF}$, and $m_i = N_{\rm E}+1$ for all $i = 1, \ldots, C$. For both $\mathcal{P}_1$ and $\mathcal{P}_2$, an additional computational cost of $\mathcal{O}(N_{\rm RF}(N_{\rm E}+1)^3)$ arises due to Singular Value Decomposition (SVD) after the optimization. In contrast, the CFS approach has a reduced complexity of $\mathcal{O}(N_{\rm RF}(N_{\rm E}+1)^3)$ solely for the SVD.

\begin{comment}
\begin{figure}
        %\hspace{1mm}
        \centering
        \includegraphics[scale=0.5]{Figures/Beamfocus.pdf}
        \caption{\small{Beamforming capability of the proposed dual-functional RX DMA design with the CFS solution, considering a panel featuring $N_{\rm RF}=8$ microstrips, each integrating $N_{\rm E} = 64$ phase-tunable metamaterials, and $P_{\rm max}=0$.}}
        \label{fig: Beamfocus}
\end{figure}    
\end{comment}

\section{Numerical Results and Discussion}
In this section, we numerically evaluate the communication and sensing performance of the proposed dual-functional RX DMA design. We have simulated a narrowband setup with a frequency of operation at $20$ GHz, where coherent channel blocks spanned $T=100$ transmissions. For the estimation of the radar targets within the AoI of the RX DMA, we have used the modified multiple signal classification algorithm presented in~\cite{FD_HMIMO_2023}. The AoI was defined at the fixed elevation angle of $\theta = 30^\circ$, azimuth angle of $\varphi \in [10^\circ, 80^\circ]$, and range of $r\in[2, 10]$ (i.e., within both the Fresnel and the Fraunhofer regions for the considered DMA panel dimensions). We have considered $K=5$ radar targets, whom $U=2$ are the uplink data-transmitting UEs, and an RX DMA equipped with $N_{\rm RF}=8$ RX RF chains, each employing $N_{\rm E}=64$ metamaterials; the former had inter-element spacing of $\lambda/2$ and the latter of $\lambda/5$. We have used $500$ Monte Carlo trials for all our simulation results, and have set AWGN's variance as $\sigma^2 = -100$ dBm as well as the UE SNR thresholds as $\gamma_u=30$ dB $\forall u$, unless otherwise stated. Lastly, for the intended optimization, we have considered a discretization of the AoI into $|\mathbb{A}|=8$ equidistant points.

In Fig.~\ref{fig: RMSE}, we evaluate the sensing performance of the proposed dual-functional RX DMA design with all three presented solutions in Section~\ref{Sec: 3} for different levels of transmit power, and compare it with the recent DMA-based RX design in~\cite{gavras2025near} that has an optimization objective the minimization of the CRB of the targets' positions. For this benchmark, we have assumed the idealized case of full knowledge of all target positions. As shown in Fig.~\ref{fig: RMSE}, and as expected, increasing the transmit SNR leads to improved target position accuracy, with the Root Mean Square Error (RMSE) converging to the corresponding PEB. As also expected, direct CRB optimization via $\mathcal{P}_1$ achieves the best sensing performance, while the alternative solutions that are based on the CRB's lower bound offer comparable results at a significantly lower computational cost. Notably, all proposed solutions herein outperform the optimization approach in \cite{gavras2025near}, even when full target position information is available. This behavior underscores the advantages of the proposed AoI-wide sensing via our area-wide PEB, allowing us to detect all targets without directly optimizing with respect to a priori knowledge of the radar targets' exact positions. 

In Fig.~\ref{fig: BF}, we investigate the BF capability of the proposed dual-functional RX DMA design with the CFS solution when the transmit power is $P_{\rm{\max}}=0$ dBm, depicting the normalized beam focusing/steering gain within the AoI. It can be observed that, for the targets lying closer to the RX panel -in fact, within its near-field region-, beam focusing is achieved around their actual position, whereas as we traverse into the far-field region, beam steering is realized towards the actual target position. It is important to recall that the proposed design prioritizes minimizing the AoI-wide PEB for sensing. This observed BF behavior suggests that, while the design does not explicitly focus on the exact target positions, it naturally enables beam focusing and steering, facilitating range and angle estimation. However, as targets move into the far field, increased range ambiguity makes this estimation progressively more challenging. Notably, this transition in BF behavior occurs within approximately one-tenth of the near-field region~\cite{liu2023near}. Lastly, it is evident from the figure that the peaks corresponding to passive objects are more spread out compared to those of the UEs. This is because passive object positions are estimated from the attenuated reflections of the UE signals, whereas UE positions can be determined directly from the LoS UL signal received by the RX.

Finally, Fig.~\ref{fig: SNR} illustrates the sensing performance of the proposed optimization solutions as a function of the common uplink instantaneous SNR threshold for the UEs, while considering a transmit power of $P_{\rm \max}=0$ dBm. It can be observed that, as $\gamma_u$ increases, more resources are allocated to communications, leading to a decline in sensing performance. It is also shown that the solutions for $\mathcal{P}_1$ and $\mathcal{P}_2$ exhibit a similar rate of degradation in sensing accuracy. In contrast, the CFS approach focuses on meeting the SNR constraints, while maintaining a constant PEB. This trend highlights the need for more sophisticated ISAC schemes as communications SNR constraints become more stringent.

\section{Conclusion}
In this paper, an RX DMA design for optimizing sensing performance within an AoI, while meeting individual QoS constraints for multi-user uplink communications, was presented. First, the area-wide CRB for the estimation of both active and passive radar targets lying in the AoI was derived, which was then used as the minimization objective for designing the DMA's analog RX BF weights under SNR threshold constraints. By leveraging the partially-connected analog BF architecture of DMAs, we transformed the proposed BF design problem into a convex formulation. A solution directly minimizing the area-wide CRB was designed, along with low complexity alternatives based on a CRB lower-bound approximation. Our simulation results showcased the accuracy of our dual-functional RX DMA design, demonstrating its sensing superiority over conventional ISAC approaches, while simultaneously guaranteeing desired multi-user uplink communication performance.

%\vspace{-0.1cm}
\bibliographystyle{IEEEtran}
\bibliography{ms}
\end{document}

%% file: Definitions.tex
\renewcommand{\b}{\mathbf{b}}

\newcommand{\e}{\mathbf{e}}

\newcommand{\h}{\mathbf{h}}

\renewcommand{\k}{\mathbf{k}}

\newcommand{\m}{\mathbf{m}}
\newcommand{\n}{\mathbf{n}}

\newcommand{\q}{\mathbf{q}}

\newcommand{\s}{\mathbf{s}}
\renewcommand{\t}{\mathbf{t}}
\renewcommand{\u}{\mathbf{u}}

\newcommand{\w}{\mathbf{w}}

\newcommand{\y}{\mathbf{y}}

\newcommand{\A}{\mathbf{A}}
\newcommand{\B}{\mathbf{B}}

\renewcommand{\H}{\mathbf{H}}
\newcommand{\I}{\mathbf{I}}

\newcommand{\N}{\mathbf{N}}

\renewcommand{\P}{\mathbf{P}}
\newcommand{\Q}{\mathbf{Q}}

\newcommand{\W}{\mathbf{W}}

\newcommand{\Y}{\mathbf{Y}}

\newcommand{\Compl}{\mbox{$\mathbb{C}$}}

\renewcommand{\Re}{\mathrm{Re}}

%% file: SPAWC.bbl
\begin{thebibliography}{10}
\providecommand{\url}[1]{#1}
\csname url@samestyle\endcsname
\providecommand{\newblock}{\relax}
\providecommand{\bibinfo}[2]{#2}
\providecommand{\BIBentrySTDinterwordspacing}{\spaceskip=0pt\relax}
\providecommand{\BIBentryALTinterwordstretchfactor}{4}
\providecommand{\BIBentryALTinterwordspacing}{\spaceskip=\fontdimen2\font plus
\BIBentryALTinterwordstretchfactor\fontdimen3\font minus \fontdimen4\font\relax}
\providecommand{\BIBforeignlanguage}[2]{{%
\expandafter\ifx\csname l@#1\endcsname\relax
\typeout{** WARNING: IEEEtran.bst: No hyphenation pattern has been}%
\typeout{** loaded for the language `#1'. Using the pattern for}%
\typeout{** the default language instead.}%
\else
\language=\csname l@#1\endcsname
\fi
#2}}
\providecommand{\BIBdecl}{\relax}
\BIBdecl

\bibitem{XLMIMO_tutorial}
Z.~Wang \emph{et~al.}, ``A tutorial on extremely large-scale {MIMO} for {6G}: {F}undamentals, signal processing, and applications,'' \emph{IEEE Commun. Surveys \& Tuts.}, early access, 2024.

\bibitem{41}
T.~Gong \emph{et~al.}, ``Holographic {MIMO} communications: Theoretical foundations, enabling technologies, and future directions,'' \emph{IEEE Commun. Surveys \& Tuts.}, vol.~26, no.~1, pp. 196--257, 2024.

\bibitem{hua2024near}
H.~Hua \emph{et~al.}, ``Near-field integrated sensing and communication with extremely large-scale antenna array,'' \emph{arXiv preprint arXiv:2407.17237}, 2024.

\bibitem{6G-DISAC_mag}
E.~Calvanese~Strinati \emph{et~al.}, ``Towards distributed and intelligent integrated sensing and communications for {6G} networks,'' \emph{IEEE Wireless Commun.}, vol.~32, no.~1, pp. 60--67, 2025.

\bibitem{Shlezinger2021Dynamic}
N.~Shlezinger \emph{et~al.}, ``Dynamic metasurface antennas for {6G} extreme massive {MIMO} communications,'' \emph{IEEE Wireless Commun.}, vol.~28, no.~2, pp. 106--113, 2021.

\bibitem{10505154}
J.~Xu \emph{et~al.}, ``Near-field wideband extremely large-scale {MIMO} transmissions with holographic metasurface-based antenna arrays,'' \emph{IEEE Transactions on Wireless Communications}, vol.~23, no.~9, pp. 12\,054--12\,067, 2024.

\bibitem{zhang2022beam}
H.~Zhang \emph{et~al.}, ``Beam focusing for near-field multiuser {MIMO} communications,'' \emph{IEEE Trans. Wireless Commun.}, vol.~21, no.~9, pp. 7476--7490, 2022.

\bibitem{Nlos_DMA}
K.~Stylianopoulos \emph{et~al.}, ``Autoregressive attention neural networks for non-line-of-sight user tracking with dynamic metasurface antennas,'' in \emph{Proc. IEEE CAMSAP}, Los Sue\~{n}os, Costa Rica, 2023.

\bibitem{NF_beam_tracking}
P.~Gavriilidis and G.~C. Alexandropoulos, ``Near-field beam tracking with extremely massive dynamic metasurface antennas,'' \emph{IEEE Trans. Wireless Commun.}, early access, 2025.

\bibitem{FD_HMIMO_2023}
I.~Gavras \emph{et~al.}, ``Full duplex holographic {MIMO} for near-field integrated sensing and communications,'' in \emph{Proc. {EUSIPCO}}, Helsinki, Finland, 2023.

\bibitem{spawc2024}
I.~Gavras and G.~C. Alexandropoulos, ``Simultaneous near-field {THz} communications and sensing with full duplex metasurface transceivers,'' in \emph{Proc. IEEE SPAWC}, Lucca, Italy, 2024.

\bibitem{gavras2025near}
------, ``Near-field localization with dynamic metasurface antennas at {TH}z: {A} {CRB} minimizing approach,'' \emph{IEEE Wireless Commun. Lett.}, early access, 2025.

\bibitem{XL_MIMO_ISAC}
G.~C. Alexandropoulos and I.~Gavras, ``Extremely large full duplex {MIMO} for simultaneous downlink communications and monostatic sensing at sub-{THz} frequencies,'' \emph{arXiv preprint arXiv:2502.10693}, 2025.

\bibitem{DMA_Losses}
P.~Gavriilidis and G.~C. Alexandropoulos, ``How do microstrip losses impact near-field beam depth in dynamic metasurface antennas?'' \emph{arXiv preprint arXiv:2503.12280}, 2025.

\bibitem{FR3}
H.~Miao \emph{et~al.}, ``A survey of new mid-band/{FR3} for {6G}: Channel measurement, characterization and modeling in outdoor environment,'' \emph{arXiv preprint arXiv:2504.06727}, 2025.

\bibitem{liu2023near}
Y.~Liu \emph{et~al.}, ``Near-field communications: {A} tutorial review,'' \emph{IEEE Open J. Commun. Soc.}, vol.~4, pp. 1999--2049, 2023.

\bibitem{5G_NR_positioning}
L.~Italiano \emph{et~al.}, ``A tutorial on {5G} positioning,'' \emph{IEEE Commun. Surveys \& Tuts.}, early access, 2024.

\bibitem{kay1993fundamentals}
S.~M. Kay, \emph{Fundamentals of Statistical Signal Processing: Estimation Theory}.\hskip 1em plus 0.5em minus 0.4em\relax Prentice-Hall, Inc., 1993.

\bibitem{keskin2022optimal}
M.~F. Keskin \emph{et~al.}, ``Optimal spatial signal design for mm{W}ave positioning under imperfect synchronization,'' \emph{IEEE Trans. Veh. Technol.}, vol.~71, no.~5, pp. 5558--5563, 2022.

\bibitem{gavriilidis2024metasurface}
P.~Gavriilidis \emph{et~al.}, ``Metasurface-based receivers with $1$-bit {ADCs} for multi-user uplink communications,'' in \emph{Proc. IEEE ICASSP}, Seoul, South Korea, 2024.

\bibitem{cvx}
I.~CVX~Research, ``{CVX}: Matlab software for disciplined convex programming, version 2.0,'' \url{http://cvxr.com/cvx}, 2012.

\bibitem{luo2010semidefinite}
Z.-Q. Luo \emph{et~al.}, ``Semidefinite relaxation of quadratic optimization problems,'' \emph{IEEE Signal Process. Mag.}, vol.~27, no.~3, pp. 20--34, 2010.

\bibitem{1}
C.~Girard, C.~Joachim, and S.~Gauthier, ``The physics of the near-field,'' \emph{Reports on Progress in Physics}, vol.~63, no.~6, p. 893, 2000.

\bibitem{nemirovski2004interior}
A.~Nemirovski, ``Interior point polynomial time methods in convex programming,'' \emph{Lecture notes}, vol.~42, no.~16, pp. 3215--3224, 2004.

\end{thebibliography}
